\documentclass{SciPost}

\hypersetup{
    colorlinks,
    linkcolor={red!50!black},
    citecolor={blue!50!black},
    urlcolor={blue!80!black}
}

\usepackage[bitstream-charter]{mathdesign}
\DeclareSymbolFont{usualmathcal}{OMS}{cmsy}{m}{n}
\DeclareSymbolFontAlphabet{\mathcal}{usualmathcal}

\fancypagestyle{SPstyle}{
\fancyhf{}
\lhead{\colorbox{scipostblue}{\bf \color{white} ~SciPost Physics Proceedings }}
\rhead{{\bf \color{scipostdeepblue} ~Submission }}

\fancyfoot[C]{\textbf{\thepage}}
}

\begin{document}

\pagestyle{SPstyle}

\begin{center}{\Large \textbf{\color{scipostdeepblue}{
High Energy Neutrino Studies in the forward direction with FASER experiment at the LHC\\
}}}\end{center}

\begin{center}\textbf{
Osamu Sato\textsuperscript{$\star$} on behalf of the FASER Collaboration.
}\end{center}

\begin{center}
{\bf } Nagoya University
\\[\baselineskip]
$\star$ \href{mailto:email1}{\small sato@flab.phys.nagoya-u.ac.jp}
\end{center}

\definecolor{palegray}{gray}{0.95}
\begin{center}
\colorbox{palegray}{
  \begin{tabular}{rr}
  \begin{minipage}{0.36\textwidth}
    \includegraphics[width=55mm]{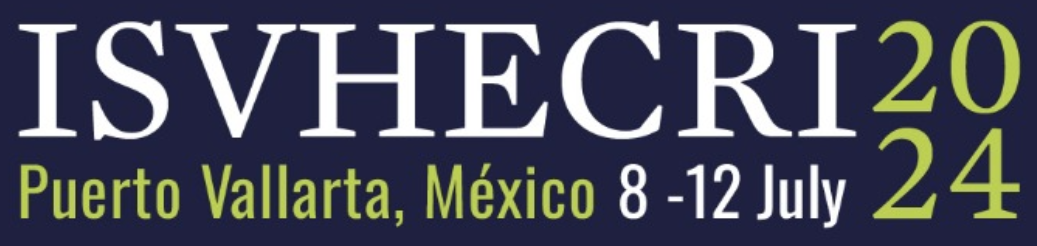}
  \end{minipage}
  &
  \begin{minipage}{0.55\textwidth}
    \begin{center} \hspace{5pt}
    {\it 22nd International Symposium on Very High \\Energy Cosmic Ray Interactions (ISVHECRI 2024)} \\
    {\it Puerto Vallarta, Mexico, 8-12 July 2024} \\
    \doi{10.21468/SciPostPhysProc.?}\\
    \end{center}
  \end{minipage}
\end{tabular}
}
\end{center}

\section*{\color{scipostdeepblue}{Abstract}}
\textbf{\boldmath{%
The FASER experiment studies the neutral decay  products from LHC collision of 13.6 TeV centre of mass energy  at 480m distant away. There  could be  Beyond Standard Model (BSM) particles such like dark photons or axion like particles etc.., and  also high energy neutrinos. The neutrino target is an Emulsion Cloud Chamber  with tungsten plates who can measure all three neutrino flavours with clear separation. Recently, FASER has measured the muon and electron neutrino cross-section with a nucleon as $ \sigma(\nu_{e} + N) = 1.2^{+0.8}_{-0.7}\times 10^{-38} cm^{2}/GeV$ and  $ \sigma(\nu_{\mu} + N) = 0.5 \pm{0.2}\times 10^{-38} cm^{2}/GeV$ in the unexplored energy region of a few hundred GeV to a few TeV so far.
}}

\vspace{\baselineskip}

\noindent\textcolor{white!90!black}{%
\fbox{\parbox{0.975\linewidth}{%
\textcolor{white!40!black}{\begin{tabular}{lr}%
  \begin{minipage}{0.6\textwidth}%
    {\small Copyright attribution to authors. \newline
    This work is a submission to SciPost Phys. Proc. \newline
    License information to appear upon publication. \newline
    Publication information to appear upon publication.}
  \end{minipage} & \begin{minipage}{0.4\textwidth}
    {\small Received Date \newline Accepted Date \newline Published Date}%
  \end{minipage}
\end{tabular}}
}}
}


\vspace{10pt}
\noindent\rule{\textwidth}{1pt}
\tableofcontents
\noindent\rule{\textwidth}{1pt}
\vspace{10pt}


\section{Introduction}
\label{sec:intro}

 K/$\pi$ meson production rate in high energy proton on nucleus interactions is an important parameter to understand / solve for too high rate of  high energy cosmic ray muon  in a few EeV energy region \cite{Muon_Puzzle}.  
Also charm hadron production rate is a key for high energy neutrino study such as  astrophysical neutrino observation.

FASER experiment aims to study neutral particles produced at LHC collision point or from decay products of collision products.
There are two main physics targets, one is neutral particle in BSM and the other is high energy neutrinos where no experimental data exist.
Here, we focussing on high energy neutrino study in FASER experiment.

\section{FASER experiment}
\subsection{FASER detector}

Size of the FASER detector is very compact because focusing on forward direction with the beam proton axis.
The detector have two parts, one is neutrino target on front and the other is electronic detector with a set of air core magnet spectrometers which act also as decay volume for BSM particles and followed by calorimeter.  The crosssection is less than 30cm $\times$ 30cm and the total length is about 7 m.

The neutrino detector is an Emulsion Cloud Chamber (ECC), a stack of 730 sets of 1 mm tungsten plates and 0.3 mm nuclear emulsion films alternatively, the cross section of 25cm $\times$ 30 cm and total weight of 1.1 tons and 1 m long.  Neutrino interaction target is mainly tungsten plates and interacted product charged tracks are recorded by emulsion plates with a sub micron spacial and mrad angular resolution by each emulsion plate. A track position and angle are measured as 3 dimensional vector and two track separation for a few $\mu$m distant tracks can be done without any ambiguity.  Tracks and interaction vertices are reconstructed using the track segments measured in  emulsion plates. A short lived particle like tau particles or charm particles can be identified through detection of kink trajectory in a short range $\sim$ 1 cm. Tungsten plates will  act also radiator for electron, its radiation length is 3.5 mm and an  electro magnetic shower can grow in several or several tenth plate depth. 
In neutrino Charge Current interactions (CC), a charged lepton is produced at the interaction point. 3 type of neutrino produced electron, mu and tau in CC.
Electron will make a electro magnetic shower, muon will penetrate without any interaction or scattering, tau will decay in a short range. 
All these features can be recognized by reconstructed tracks in emulsion, so 3 neutrino flavour will be categorized as they are.
A large value of tungsten radiation length, $X_{0}$ =3.5mm, will act as also scattering material for charged particles. The Multiple Coulomb Scattering of charged tracks at penetrating neutrino target will give us their momentum information. And the momentum values obtained by tracking in the ECC are used for event selection to separate from background or charged lepton, so neutrino flavour identification.
  
The electronic detector has mainly 3 components,  3 sets of  about 1 m long with 10cm $\phi$ air core di-pole magnets, silicon stripe detectors for tracking before and after the magnets and measure curvature and its momentum, calorimeter .
A uniform 0.55 T magnetic field perpendicular to line of sight and thanks to tracking by fine pitch (80$\mu$m) silicon stripe detector planes, the momentum measurement accuracy  $\sim$ 7.5\% (at 100 GeV/c)  to 15\% (at 1 TeV/c) .
  
\begin{figure}[h]
\begin{minipage}{34pc}
\includegraphics[width=27pc]{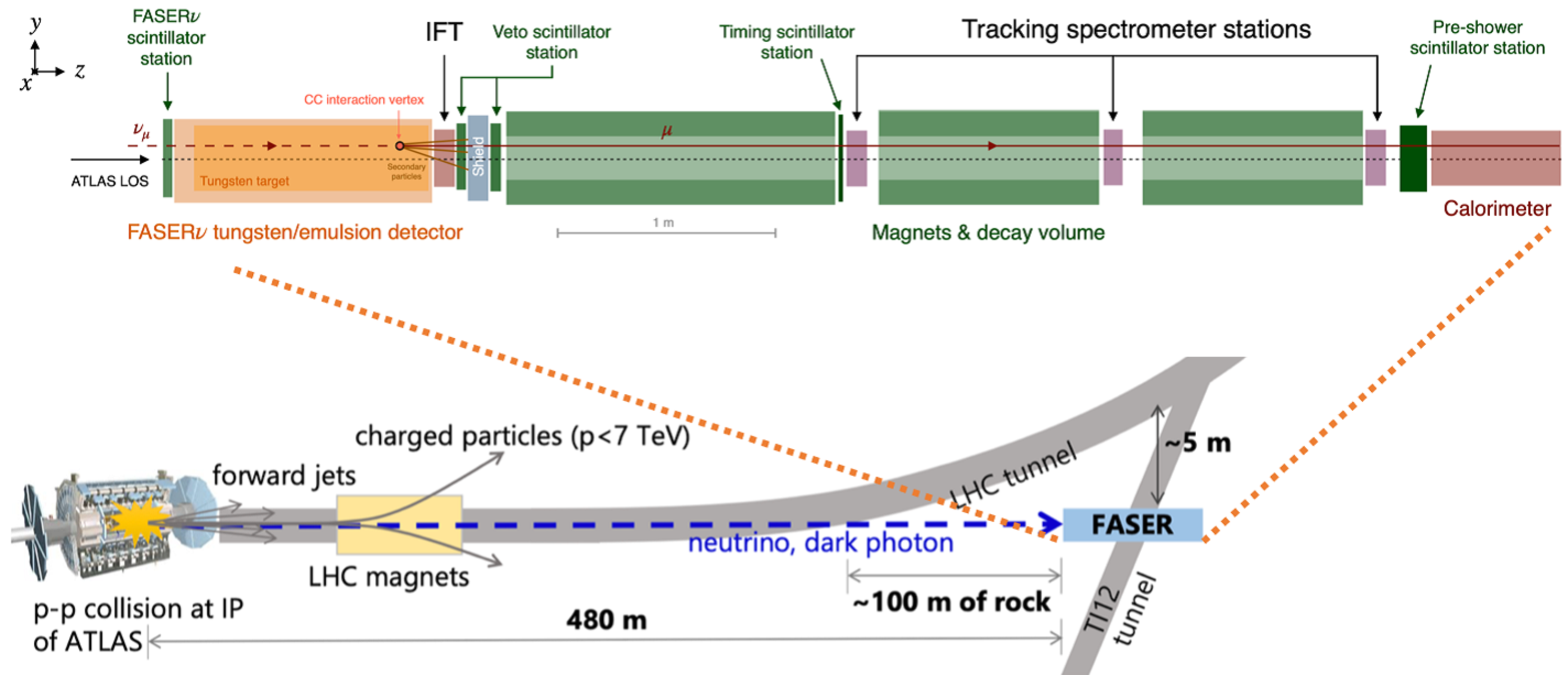}
\caption{\label{FASER} The location and the schematic view of the FASER detector.}
\end{minipage} 
\end{figure}

\subsection{Beam acquisition}

FASER collecting data from LHC run3, 2022.
93.1 fb$^{-1} $beam were delivered and 90.7 fb$^{-1}$ recorded till June  2024.
The neutrino targets (FASER$\nu$ box hereafter) were exchanged 3 times a year to keep the accumulated track density in emulsion films less than $10^{6}$ /cm$^{2}$.
6 neutrino targets were irradiated till June2024. 
In next section, a part of data from second FASER$\nu$ box analysis results will be described. 
  

\section{Analysis of neutrino interaction }
\label{analysis}

\subsection{Neutrino search by Electronic detectors}

Events from a good quality data in 2022 run were used for the analysis.
Treating FASER$\nu$ box as a just a  neutrino target and requesting a good track consistent with $\mu$ from $\mu$neutrino CC are reconstructed in spectrometers.
The track should be in fiducial area where radius from  air core centre $<$ 95mm and momentum  reconstructed more than 100 GeV/c and its slope at origin is less than 25 mrad, And extrapolate position in front veto plane should be consistent with FASER$\nu$ target area (r $<$ 120mm) .
The event expectation from GENIE simulation is 151 $\pm$ 41 with background 0.08$ \pm$ 1.83 for the given beam and selection condition.

Observed event was 153$^{+13} _{-12}$ events  by unblind the data.
This corresponds to 16 $\sigma$ signal significance, and first direct observation of collider neutrinos.
Muon signs are measured and both muon neutrinos and anti muon neutrinos  are detected.
The muon momentum distribution is also measured.
More detail can be found in \cite{electronic_neutrino}.

\begin{figure}[h]
\begin{minipage}{34pc}
\includegraphics[width=28pc]{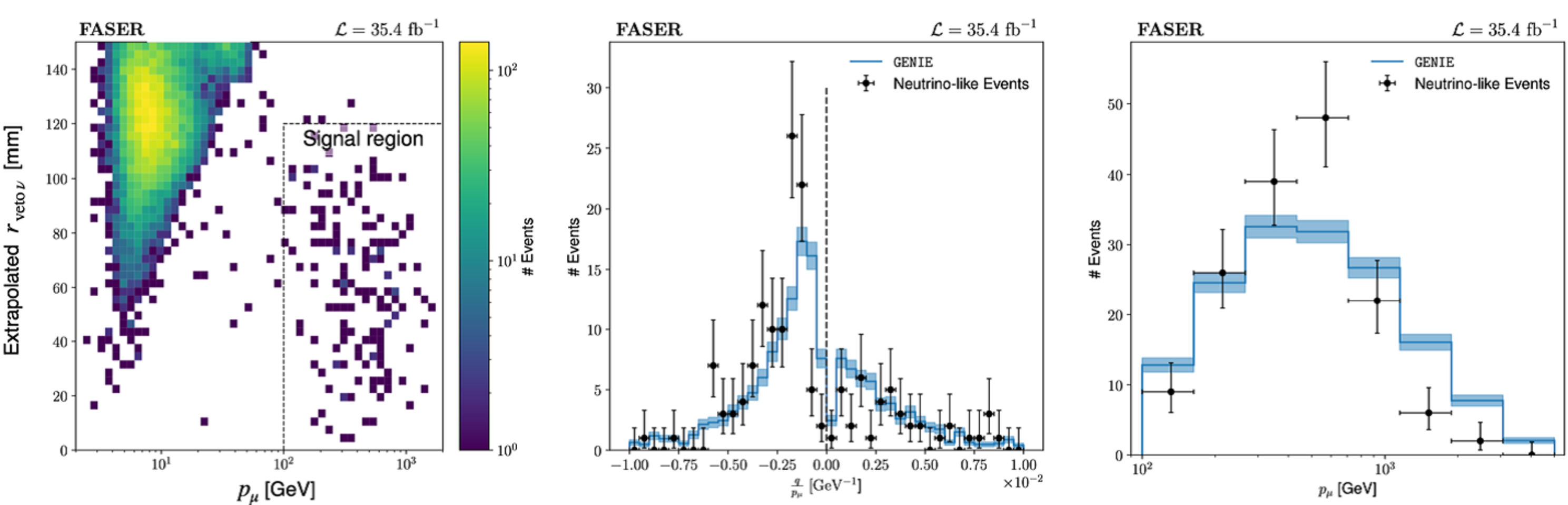}
\caption{\label{eldet} Detected position, Inverse momentum $\times$ charge and momentum distribution of the neutrinos detected by electronic detector.}
\end{minipage} 
\end{figure}

\subsection{Neutrino search by Emulsion Cloud Chamber}

Nuclear Emulsion films are produced \cite{emulsion} at Nagoya University  and are sent to CERN to assembling with tungsten plates.
A FASER$\nu$ box is an stack of 730 sets of emulsion film and tungsten film alternatively, with vacuum packed each 10 film and tungsten plates. 
It is mounted on front of FASER electronic detector complex.
During the beam exposure the temperature stability around FASER$\nu$ box is kept $\pm$ 0.1 degree to avoid alignment destruction inside the vacuum pack due to the difference of thermal expansion coefficient.   
After beam exposure the FASER$\nu$ box are extracted and de-packing the emulsion films at CERN dark room and films are chemical developed.
The developed films are cleaned and sent to Nagoya University to scan track information by automatic high speed scanning system, HTS \cite{HTS}.
After reading out tracks information from each emulsion film, the tracks and vertices are reconstructed with very nice alignment accuracy of $\sim$ 0.5um.

Energy determination for muon like charged particles, i.e. not electro magnetic particles, are based on measurement of displacement due to Multiple Coulomb Scattering.
Charged track momentum can be measured up to a few TeV by the resolution of dP/P ~30\% at 200 GeV/c and <50\% for higher momentum of interest thanks to fine spacial resolution and good alignment of detector module by high track density,
The performances of the momentum measurement are proven with given momentum test beam in 2023.
The ,momentum of 300 GeV/c muon beam are measured as 286 GeV/c and width of the distribution corresponding to dP/P ~30\% consistent with Mote Carlo Simulation.  
For energy determination on electron like particle case, number of track segment in an electro magnetic shower maximum grow depth $\pm$ 3 emulsion films.
The energy resolution dE/E of an electron is 25\% at 200 GeV and 25-40\% for higher energy of interest.   

For this time a sub-sample of data from the second FASER$\nu$ box (9.5fb$^{-1}$)  in 2022 were analysed for neutrino nucleon interaction cross section measurement..
A fiducial volume are set to 128.6kg out of a  FASER{$\nu$} 1.1 tons target module.
A set of criteria was set to enhance CC events and reject neutral hadron interactions..
Number of tracks forming the interacted vertex candidate should have more than 4 tracks and charged lepton candidate should have momentum bigger than 200 GeV/c and tan $\theta$  is larger than 0.005, angle between charged lepton and hadron axis should be bigger than 90 degree.

\begin{figure}[h]
\begin{minipage}{34pc}
\includegraphics[width=22pc]{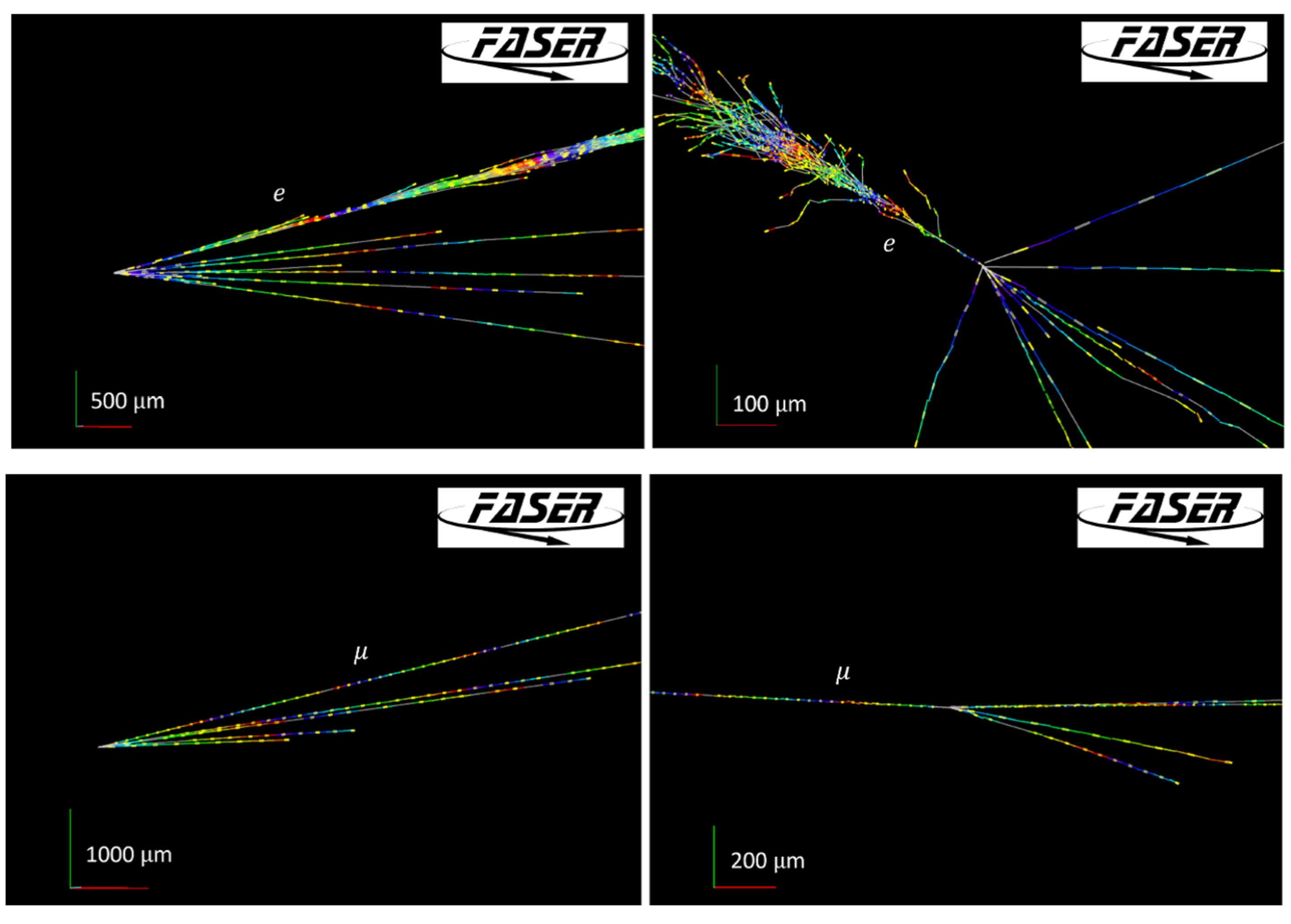}
\caption{\label{display} Event Display of CC Interaction candidates. Top: An electron neutrino Bottom: A muon neutrino candidate, Left: beam direction along the horizontal axis. Right: beam axis perpendicular to the paper. }
\end{minipage} 
\end{figure}

4 electron neutrino CC candidates were observed under 1.1 -3.3 signal expectation with background 0.025$^{+0.012} _{-0.010}$.
This is the first observation of electron neutrino produced at the LHC. 
8 $\mu$ neutrino CC candidates were observed under 6.5-12.4 signal expectation with background 0.22$^{+0.09}_ {-0.07}$.
The event display of a muon neutrino CC and an electron neutrino CC candidates is also shown in Figure.\ref{display}

\begin{figure}[h]
\begin{minipage}{34pc}
\includegraphics[width=25pc]{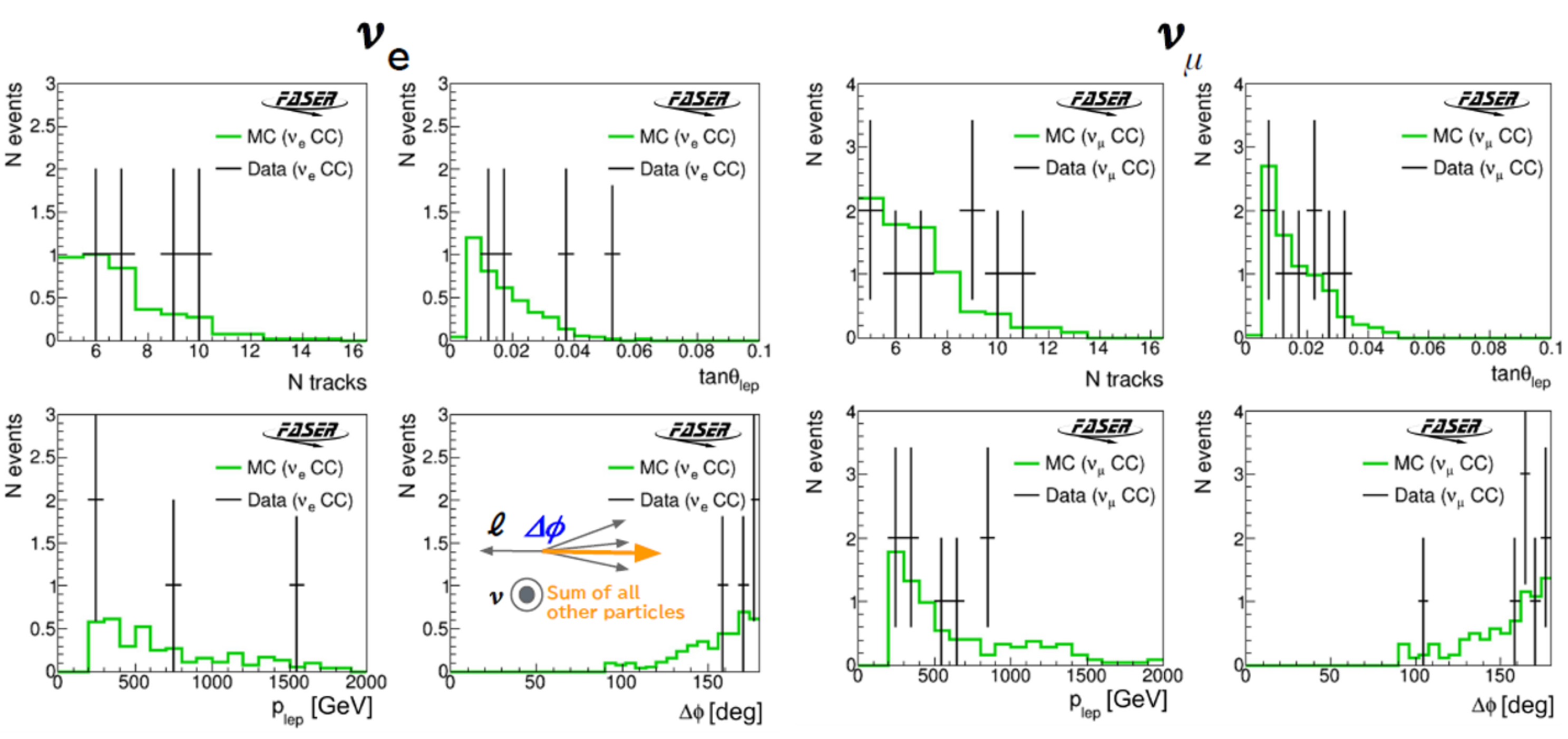}
\caption{\label{keypar} Distributions of key parameters of the detected $\nu_{e}$ (Left) and $\nu_{\mu}$ (Right). Number of charged tracks at the vertex, The charged lepton emission angle, The charged lepton momentum, Angle between hadrons and the charged lepton in perpendicular to the beam. }
\end{minipage} 
\end{figure}

\begin{figure}[h]
\begin{minipage}{34pc}
\includegraphics[width=25pc]{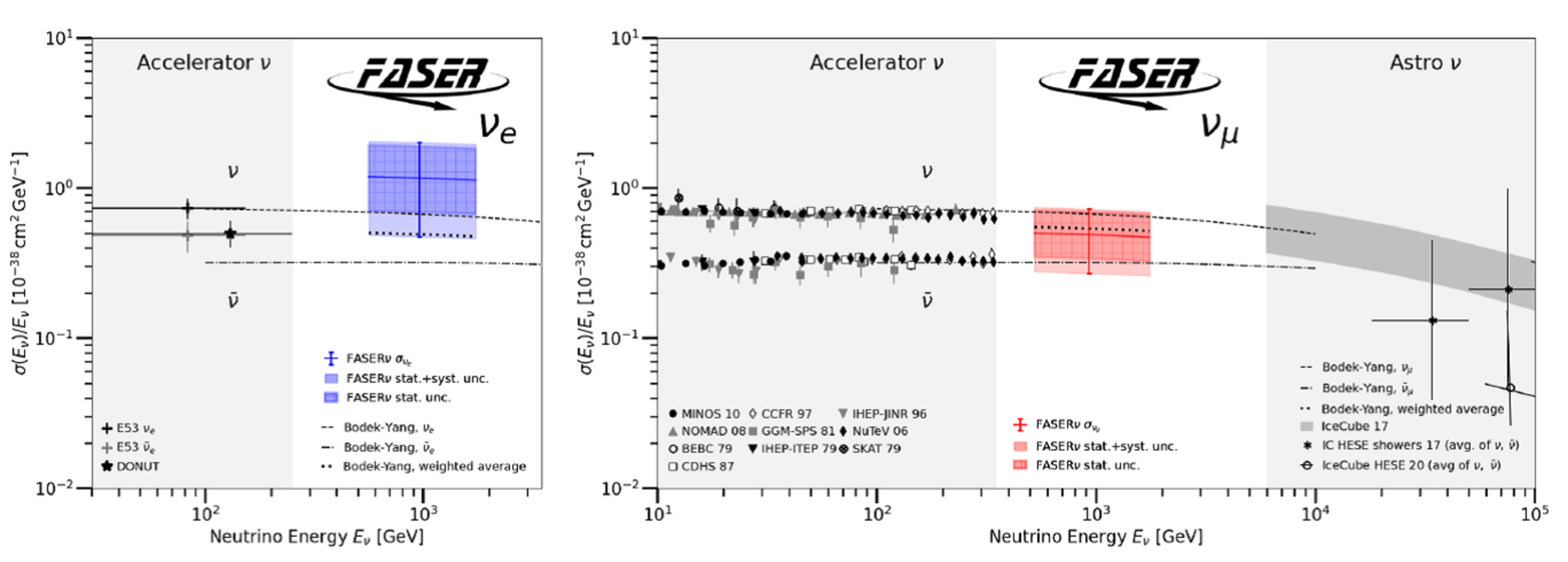}
\caption{\label{cross_section} Measured Neutrino-Nucleon cross section blue $(\nu_{e}$) and red ($\nu_{\mu}$) .}
\end{minipage} 
\end{figure}

Distribution of the key parameters used in the selection criteria is shown in Figure.\ref{keypar}.
The neutrino cross section using the observed CC events are measured as \\
$ \sigma(\nu_{e} + N) = 1.2^{+0.8}_{-0.7}\times 10^{-38} cm^{2}/GeV$ and  $ \sigma(\nu_{\mu} + N) = 0.5 \pm{0.2}\times 10^{-38} cm^{2}/GeV$. \cite{first_cross}, Figure.\ref{cross_section}

\section{Prospects}

Here $\nu_{e}$ and $\nu_{\mu}$ cross section are reported using 1.7 \% of accumulated data of FASER$\nu$ so far. 
The data accumulation will continue to the at the end of LHC-RUN3 (up to 2025 )
In total about 10,000 neutrino CC neutrino will be collected.

A few tenth  tau neutrino CC will be observed and the tau neutrino cross section will be measured.
So that the neutrino nucleon interaction cross section of all 3 neutrino flavours will be measured. 
A lepton flavour universality check on interaction cross-section can be performed in neutral lepton sector.
One can extract information on forward meson production rate (Charm/K/p) at LHC collision point from the differential cross-section distribution of 3 neutrino flavour on neutrino energy since mother hadrons composition of each neutrino flavour is different \cite{Neutrino_Flux}.

After LHC run3, a new Forward Physics Facility \cite{FPF}  is under discussion.
Where the target mass of FASER$\nu$ boxes is 20 tons and large cross section (40cm $\times$ 40cm)  and longer (8 m)  neutrino active target detector followed by electronic detector is considered to detect a million neutrino CC and  2300(SIBYLL) -20000(DPMJET) tau neutrino CC observation is expected.

\section{Conclusion}

FASER$\nu$ is a project analysing high energy, unexplored so far, neutrinos coming from LHC collision products.
FASER$\nu$ studies three flavour neutrinos at the high energy frontier and neutral lepton flavour universality on coupling constant and extract also neutrino mother hadron produced information in the forward direction at LHC collision. 
In total a 10,000 $\nu$ CC interactions expected in LHC Run 3 (2022-2025, 250 fb$^{-1}$).

First direct observation of collider neutrino was reported by the detection of 153 $\mu$ neutrino CC interactions with FASER (signal significance of 16 $\sigma$) electronic detector. 
And here using nuclear emulsion tracking, the first measurement of neutrino-nucleon cross section at a few hundred to a few TeV energy region was reported by a subsample of FASER$\nu$ data.
8 $\mu$ neutrino  CC candidates and 4 electron neutrino  CC candidates gives measurement of the cross section as 
$ \sigma(\nu_{e} + N) = 1.2^{+0.8}_{-0.7}\times 10^{-38} cm^{2}/GeV$ and $\sigma(\nu_{\mu} + N) = 0.5 \pm{0.2}\times 10^{-38} cm^{2}/GeV $

Tau neutrino observation and cross section measurement will come soon and all 3 neutrino flavours cross section will be reported by FASER$\nu$

\section*{Acknowledgements}

I would like to thank the collaboration colleagues for the hard operation and the analysis.

\bibliography{SciPost_Example_BiBTeX_File.bib}


\end{document}